\documentclass[useAMS,usenatbib]{mn2e} \input{epsf}

\newif\ifAMStwofonts 

\usepackage{lscape}

\def\lesssim{\mathrel{\hbox{\rlap{\hbox{\lower4pt\hbox{$\sim$}}}\hbox{$<$}}}}
\def\gtrsim{\mathrel{\hbox{\rlap{\hbox{\lower4pt\hbox{$\sim$}}}\hbox{$>$}}}}

\def\msol{${\rm M}_{\odot}$}
\def\rsol{${\rm R}_{\odot}$~}

\def\l_lsun{$\log{L/\rm L_{\odot}}$~}
\def\masa_msun{$M/ \rm M_{\odot}$~}

\def\m_mstar{$M/M_{*}$~}

\title[Evolutionary model for PSR~J1311-3430]{An evolutionary model for the gamma-ray system PSR~J1311-3430 and its companion}

\author[O. G. Benvenuto, M. A. De Vito \& J. E. Horvath]
{O. G. Benvenuto$^{1,2}$\thanks{Member of  the Carrera del Investigador
Cient\'{\i}fico, Comisi\'on de  Investigaciones Cient\'{\i}ficas de la
Provincia    de    Buenos    Aires    (CIC). Email:
obenvenuto@fcaglp.unlp.edu.ar}
M.A. De Vito$^{1,2}$\thanks{Member of  the Carrera del Investigador
Cient\'{\i}fico, Consejo Nacional de Investigaciones Cient\'{\i}ficas y
T\'ecnicas (CONICET). Email: adevito@fcaglp.unlp.edu.ar},
J. E. Horvath$^{3}$\thanks{Email:foton@astro.iag.usp.br} \\
$^{1}$ Facultad de Ciencias Astron\'omicas y Geof\'{\i}sicas, Universidad
Nacional de La Plata (UNLP), \\ Paseo del Bosque S/N, B1900FWA, La Plata, Argentina\\
$^{2}$ Instituto de Astrof\'{\i}sica de La Plata, IALP, CCT-CONICET-UNLP, Argentina\\
$^{3}$ Instituto de Astronomia, Geof\'{\i}sica e Ci\^encias Atmosf\'ericas, Universidade de S\~ao Paulo\\
R. do Mat\~ao 1226 (05508-090), Cidade Universit\'aria, S\~ao Paulo SP, Brazil}

\begin{document}

\date{November 18}

\pagerange{\pageref{firstpage}--\pageref{lastpage}} \pubyear{2013}

\maketitle

\label{firstpage}

\begin{abstract} 

The most  recent member of  the millisecond pulsar with  very low-mass
companions   and   short   orbital   periods   class,   PSR~J1311-3430
\citep{2012arXiv1211.1385P} is a remarkable  object in various senses.
Besides  being  the  first  discovered  in  gamma-rays,  its  measured
features include the  very low or absent hydrogen content.  We show in
this Letter that  this important piece of information leads  to a very
restricted range of  initial periods for a given donor  mass. For that
purpose, we  calculate in  detail the evolution  of the  binary system
self-consistently,  including mass  transfer and  evaporation, finding
the  features of  the new  evolutionary path  leading to  the observed
configuration.  It  is also  important  to  remark that  the  detailed
evolutionary history  of the  system naturally leads  to a  high final
pulsar mass, as it seems to be demanded by observations.

\end{abstract}

\begin{keywords} (stars:) pulsars: general ---
(stars:) pulsars: individual (PSR~J1311-3430) ---
stars: evolution ---
(stars:) binaries (including multiple): close
\end{keywords}

\section{INTRODUCTION} \label{sec:intro}

The discovery of the $2 ms$, 93~min orbit, ``black widow'' (BW) pulsar
PSR~J1311-3430  in  gamma-rays  \citep{2012arXiv1211.1385P}  comes  to
increase  the   number  of   such  systems   in  a   very  interesting
fashion. This  is the first time  that gamma-ray pulsations led  to an
identification  without  the  need  of  low-energy  counterparts,  and
additional work  \citep{2012ApJ...760L..36R} indicated the  absence or
very low level of H in  the companion, plus the suggestive evidence of
a large mass of the pulsar itself.  These facts are quite important to
reconstruct the evolutionary path that led to the present state.

In  a  former  work  \citep{2012ApJ...753L..33B} we  have  modelled  a
similar system, namely the binary millisecond pulsar PSR~J1719-1438 in
a    2.2~h   orbit    featuring    a   very    low   mass    companion
\citep{2011Sci...333.1717B}.  We  shall show  now that  the additional
information  gathered  from PSR~J1311-3430  is  useful  to refine  the
models  and restrict  the possible  initial states  to a  rather small
range, for a given assumed value of the initial donor mass.

We  have described  the features  and tests  of the  evolutionary code
employed       in       these       studies       elsewhere       (see
\citealt{2012ApJ...753L..33B};    and    references    therein).     A
simultaneous integration of the  stellar structure, mass transfer rate
(if any) and orbit evolution is  performed taking into account i) {\it
  accretion}  onto the  neutron star  (NS); ii)  the {\it  evaporating
  wind} driven by the pulsar  radiation; and iii) the {\it irradiation
  feedback}, present when the donor  star that transfers mass onto the
NS and  accretion illuminates the  donor star modifying  its evolution
(see \citealt{2004A&A...423..281B} for further details on this item).

The three effects have been shown to be important for the evolution of
PSR~J1719-1438.   The case  of  PSR~J1311-3430  adds considerable  new
information  to  address  because spectroscopic/photometric  data  has
shown  that H  is  almost absent  \citep{2012ApJ...760L..36R}, at  the
level of a  number abundance $n_{H} < 10^{-5}$ in  contrast with other
members of the group\footnote{It corresponds  to a H mass abundance $X
  < 2.5 \times 10^{-6}$ for a helium-dominated composition.}; and also
that a high-mass for the pulsar is favoured, depending somewhat on the
interpretation of the light curve but  bounded by $M_{PSR} \geq 2.1 \,
M_{\odot}$. We shall  show now that the evolutionary  models can match
these  and the  rest of  the observed  parameters, constraining  quite
tightly the initial state of the system.

\section{CALCULATIONS} \label{sec:calcu}

The evolution  of PSR~J1311-3430 system  has been modelled  within the
same scenario depicted in  \citet{2012ApJ...753L..33B} starting with a
normal, solar composition star and  a just-formed NS orbiting together
with an initial period $P_{i} \lesssim 1$~d.

When the donor star radius $R_{2}$ equals the radius of the Roche Lobe
$R_{L}$\footnote{We shall  refer the  NS (donor  star) as  the primary
  (secondary), with subindex 1~(2).}, it occurs the onset of the Roche
Lobe  Overflow  (RLOF). Around  this  epoch  tidal dissipation  forces
quickly circularise the  orbit.  The donor star  transfers mass across
the  Lagrangian point  $L_{1}$ towards  the NS,  causing the  orbit to
evolve.  As  in \citet{2002ApJ...565.1107P}, we have  parametrized the
uncertain  fraction of  mass effectively  accreted  by the  NS with  a
quantity $\beta > 0$ (that is,  ${\dot M}_{1} = -\beta {\dot M}_{2}$),
assuming that it is always below the Eddington limit ${\dot M}_{Edd} =
2 \times 10^{-8}\;  M_{\odot}\; yr^{-1}$. Since, in  general, $\beta <
1$  some material  is lost  from  the system,  carrying away  specific
angular momentum of  the secondary. Fortunately, the  value of $\beta$
is not critical  in determining the evolution of this  kind of systems
\citep{2012MNRAS.421.2206D}, and  so we have assumed  an average value
of  $\beta=1/2$.  Gravitational  radiation \citep{1971ctf..book.....L}
and magnetic braking \citep{1981A&A...100L...7V}  are known to provide
relevant angular  momentum sinks. Additionally, the  $\dot{M}_{2}$ due
to   RLOF   has  been   described   by   the  expressions   given   in
\citet{1988A&A...202...93R}.

As   stated   in    \citet{2012ApJ...753L..33B},   this   ``standard''
prescription  valid  for  LMXBs  \citep{2002ApJ...565.1107P}  must  be
supplemented with a wind  evaporation $\dot{M}_{2,evap}$ law. Here, we
shall  assume the  expressions  given by  \citet{1992MNRAS.254P..19S},
namely

\begin{equation}
{\dot M}_{2,evap} = - {f \over{2 v_{2,esc}^{2}}} L_{P} {\biggl(
{R_{2} \over{a}} \biggr)}^{2} ,
\end{equation}

where  the spin  down luminosity  $L_{P}$ of  the pulsar  is given  by
$L_{P}= 4 \pi^{2} I_{1} P_{1} {\dot  P}_{1}$ ($I_{1}$ is the moment of
inertia  of the  NS,  $P_{1}$ is  its period  and  ${\dot P}_{1}$  its
spin-down rate),  $v_{2,esc}$ is  the escape  velocity from  the donor
star surface, $a$ is the semi-axis of the circular orbit and $f$ is an
efficiency  factor. As  in \citet{2012ApJ...753L..33B},  we shall  set
$f\;  L_{P}=  0.04  L_{\odot}$.    While  irradiation  feedback  is  a
fundamental  ingredient in  computing the  mass transfer  evolution at
early stages, we  have found that it does not  affect the evolutionary
path  of the  system  in the  mass-orbital period  plane.   As we  are
interested here on  the final evolution of the system,  in this Letter
we shall neglect irradiation feedback.

The  evolution of  the  system  has been  computed  with our  detailed
(Henyey)  evolutionary code  described in  \citet{2003MNRAS.342...50B}
and  \citet{2012MNRAS.421.2206D}.  We  started with  a fiducial  donor
mass of $2\; M_{\odot}$ together with  a $1.4~M_{\odot}$ NS, as in the
case of PSR~J1719-1438.  The range of initial donor  masses is bounded
from below  by the  isolated-star evolution timescale  ($M_{2} \approx
1.0~M_{\odot}$), which should  be short enough to allow  for the onset
of  mass  transfer; and  from  above  by $M_{2}  \geq  3.5~M_{\odot}$,
because  above  that  initial  value the  mass  transfer  is  unstable
\citep{2002ApJ...565.1107P}.  As  in  the  case of  the  evolution  of
PSR~J1719-1438, $P_{i}$ must be  very short.  However, extremely short
$P_{i}$ would cause the mass transfer  to start at the ZAMS; while too
long $P_{i}$  values would render  a detached  wide orbit after  a few
$Gyr$, not the observed BW-type system.

Even  if these  conditions are  essentially the  same as  the ones  we
studied   for  PSR~J1719-1438,   the   strong  upper   limit  set   by
\citet{2012ApJ...760L..36R}  on  the   H  abundance  allows  an
important refinement of  the models. Its solution rests  on a delicate
interplay between the  dynamical evolution of the  close binary system
(CBS)  and  the structure  of  the  donor  star  in a  novel  fashion,
described as follows.

If  the system  started  at low  periods,  around $P_{i}=0.7$~d,  mass
transfer  starts  when  H  core  abundance  is  $X_{c}^{RLOF}  \approx
0.39$.  From this  initial condition  on, $T_{c}$  drops slowing  down
H-burning.  When $M_{2}  \approx 0.053\;  M_{\odot}$ the  star becomes
completely  convective  making it  chemically  homogeneous  up to  the
photosphere  with $X= 0.099$.   Since for  a larger  $P_{i}$ the
onset of the RLOF occurs later, $X_{c}^{RLOF}$ will have a lower value
and $X$ gets smaller. Eventually, there is a $P_{i}$ value
$P_{i}\approx  0.86$~d  for  which  $X_{c}^{RLOF}=0$,  this  condition
holding for every period longer than that.

For  this  kind of  CBSs,  there  exists  a bifurcation  period  value
$P_{b}$: if $P_{i}  > P_{b}$ the CBS evolves to  an open configuration
forming a low-mass helium white dwarf,  while $P_{i} < P_{b}$ leads to
the kind of BW systems we are interested in.  For $P_{i} < P_{b}$, the
minimum  orbital  period  attained  during  evolution  decreases  when
$P_{i}$ increases\footnote{This is due to the fact that the larger the
  $P_{i}$ the star has a higher  mean molecular weight $\mu$. The star
  attains  a  partially degenerate  interior  at  larger densities,  a
  smaller   Roche   lobe   and    consequently   a   shorter   orbital
  period. Partial degeneracy is a necessary condition for the orbit of
  the    system   to    start    to   evolve    to   larger    periods
  \citep{1986A&A...155...51S}.}.   This  behaviour   can  be  seen  in
Fig.~\ref{Fig:periodos},  where for  the  system  considered here  the
bifurcation period is between 0.880 and 0.890~d.

For  $P_{i}   \lesssim  P_{b}$   there  occur  conditions   for  which
$X_{c}^{RLOF}=0$ and  mass transfer is  able to remove most  of H-rich
outer  layers.   The donor  star  becomes  a  very low-mass  He  star.
Remarkably, such configuration  leads to masses and  orbital period in
excellent   agreement   with   observations.   This   tight   interval
0.87~$\lesssim  P_{i}/d  \leq$~0.90  for PSR~J1311-3430  is  the  only
interval  accepted  in these  calculations,  a  shorter $P_{i}$  would
produce  H-rich  BW  systems.  These  features  are  clearly  seen  in
Fig.~\ref{Fig:abundancias}.

We  would like  to emphasise  that a  large value  of the  pulsar mass
\citep{2012ApJ...760L..36R}  is a  natural  outcome  of the  evolution
calculations, here  restricted to  a reasonable but  uncertain initial
value set to  $M_{1} = 1.4~M_{\odot}$ for  definiteness. The evolution
of  the   NS  (pulsar)   and  donor  star   masses  are   depicted  in
Fig.~\ref{Fig:masas}.  The final value of the NS mass is not sensitive
to $P_{i}$ as expected, and reflects the integrated accretion over the
system lifetime.  While values $\geq 2.2~M_{\odot}$  result, it should
be kept in mind that  the accretion efficiency parametrised by $\beta$
was held fixed at 0.5 in  all the calculation. A detailed modelling of
the accretion could refine this rough time-average and eventually push
the mass  to higher values  without any serious conflict  with present
observations. However, the problem of  explaining the presence of very
massive NS in these systems would call \citep{2012ApJ...760L..36R} for
a    major    revision    of     current    NS    core    microphysics
(\citealt{2006ennp.conf..188H}; \citealt{2007PhR...442..109L}).

In  Fig.~\ref{Fig:periodos_t} we  show  the period  evolution for  the
considered      CBSs.      This      figure,     complementary      of
Fig.~\ref{Fig:periodos},  provides  the  time scale  involved  in  the
evolution of the  CBSs. Finally, in Fig.~\ref{Fig:radios}  we show the
evolution of the radius of the  donor star and its corresponding Roche
lobe   for  $P_{i}=0.88$~d,   which  represents   a  good   model  for
PSR~J1311-3430.

In  Table~\ref{Tabla:datos} we  present some  relevant quantities  for
each of  the computed models. From  that, we see that  all models with
$P_{i}   <   P_{b}$  reproduce   the   main   characteristics  of   BW
systems. Particularly, the system containing the pulsar PSR~J1311-3430
is best represented by models with $P_i = 0.870 - 0.880$~d, just below
the bifurcation value.

Let us briefly address the effect  of the present uncertainties in the
ingredients  of the  model.  The  value of  $P_{i}$  favoured for  the
occurrence  of PSR  J1311-3430 is  dependent  on the  strength of  the
magnetic braking.   We have  tested it by  introducing a  factor lower
than one  in the standard expression.  If the strength is  half of the
standard  one,  the  whole  scenario  is  essentially  unchanged;  the
favoured  period  is  0.76~d  and  companions to  BWs  without  H  are
possible. However if  magnetic braking were four  times less efficient
it is still  possible to find systems with the  observed period but in
all cases the companion  retains some H. Thus, if so,  a model for PSR
J1311-3430  would  be  not  possible.   Regarding  the  value  of  the
parameter $f\; L_{P}$,  work in progress shows that if  it were larger
than assumed here, the qualitative behaviour of the evolution would be
essentially the same, but the final evolution to large orbital periods
would be markedly faster posing no problem for the whole model.

Performing a  detailed calculation  of the  probability of  having BWs
without (or  trace) H is beyond  the scope of this  paper. However, in
order to make  some preliminary estimations, we  have computed several
models for  the range of periods  that evolve to BW  configuration. We
found that  for a  standard magnetic braking,  systems evolving  to BW
have $ 0.500 \leq P_{i}/d \leq  0.887$; being $0.873 \leq P_{i}/d \leq
0.887$  for   $X<10^{-5}$  ($0.865   \leq  P_{i}/d  \leq   0.887$  for
$X<10^{-3}$). Meanwhile, for the case of magnetic braking with half of
the standard strength,  the range of $P_i$ evolving to  BWs is $ 0.500
\leq P_{i}/d  \leq 0.769$; being  $0.760 \leq P_{i}/d \leq  0.769$ for
$X<10^{-5}$ ($0.753 \leq P_{i}/d \leq  0.769$ for $X<10^{-3}$). So, if
the initial distribution  of $P_i$ were uniform, we would  have 6\% of
the  BWs with  $X<10^{-3}$  and  3\% with  $X<10^{-5}$  for both,  the
standard magnetic braking strength and a half of it.

\section{CONCLUSIONS}

A new  path for  CBSs evolving into  millisecond pulsar-very  low mass
companion final  states has been  studied and  applied to the  case of
PSR~J1719-1438  in \citet{2012ApJ...753L..33B}.   In  this Letter,  we
have applied  the same ideas to  the latest member just  discovered in
gamma-rays    \citep{2012arXiv1211.1385P},     namely    the    system
PSR~J1311-3430.   These  self-consistent  calculations  including  all
three essential  ingredients (accretion, winds and  illumination) show
that theory  is able to account  for the formation of  BW systems with
He-dominated donor composition on a few $Gyr$ timescale.

\begin{figure}
\epsfysize=300pt
\epsfbox{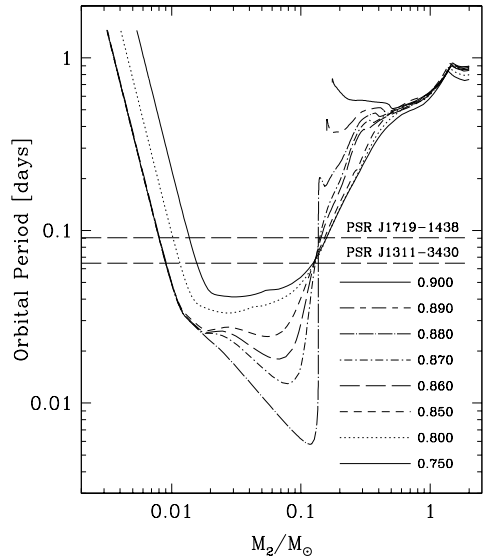}
\caption{Orbital  period-mass relation  for the  $2 M_{\odot}$  normal
  donor star evolving together with a $1.4 M_{\odot}$ neutron star for
  different  values of  the initial  periods (in  days). The  observed
  orbital  periods of  PSR~J1719-1438 and  PSR~J1311-3430 are  denoted
  with dashed lines. \label{Fig:periodos}}
\end{figure}

\begin{figure}
\epsfysize=300pt
\epsfbox{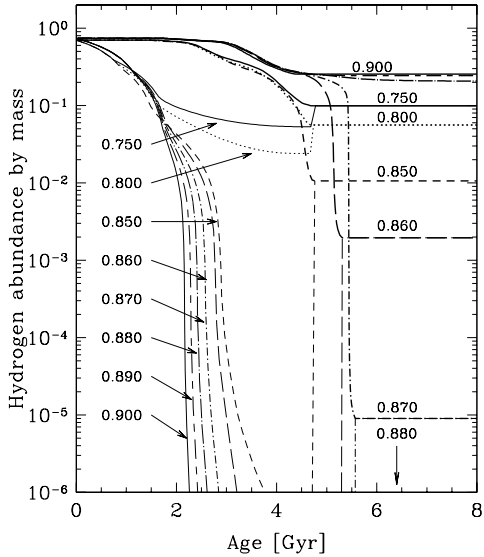}
\caption{The evolution of the H  abundance of the donor star at
  its center  (thin lines)  and surface (thick  lines) for  the models
  included in Fig.~\ref{Fig:periodos}.  Depending  on the value of the
  initial period,  at a  given age the  star becomes  fully convective
  being chemically homogeneous. \label{Fig:abundancias} }
\end{figure}

\begin{figure}
\epsfysize=300pt
\epsfbox{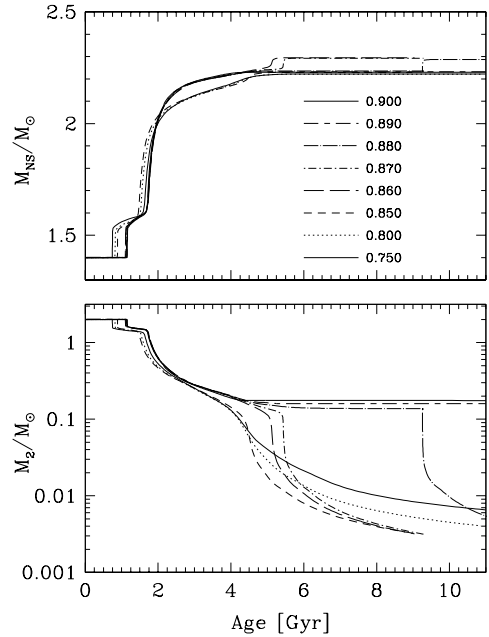} 
\caption{The evolution of  the masses of the donor  star (lower panel)
  and  neutron star  (upper panel)  for the  same systems  included in
  Fig.~\ref{Fig:periodos}.  The labels  of  the  different line  types
  denote     the    values     of    the     initial    period     (in
  days). \label{Fig:masas}}
\end{figure}

\begin{figure}
\epsfysize=300pt
\epsfbox{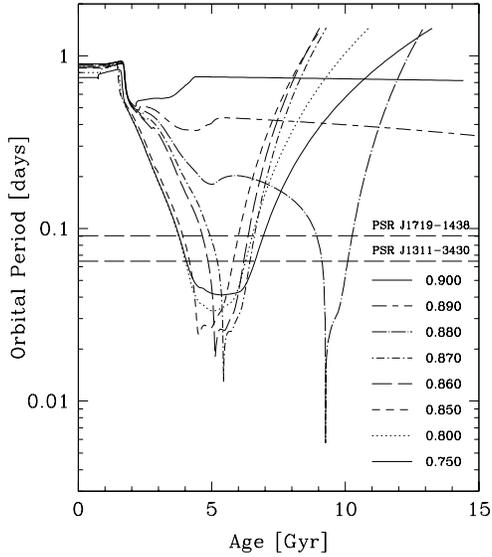} 
\caption{Orbital  period evolution  for the  same systems  included in
  Fig.~\ref{Fig:periodos}
\label{Fig:periodos_t}}
\end{figure}

\begin{figure}
\epsfysize=300pt \epsfbox{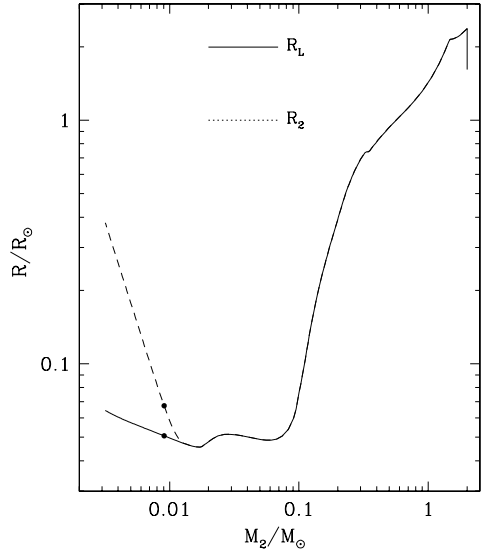} 
\caption{The radius  of the donor  star $R_{2}$ (solid lines)  and its
  corresponding Roche lobe  $R_{L}$ (dashed lines) for the  case of an
  initial period of 0.88~days.  Points  denote the conditions at which
  the system  attains the  orbital period  of PSR~J1311-3430  with low
  mass in detached conditions. \label{Fig:radios} }
\end{figure}

The initial conditions  for the particular case  of PSR~J1311-3430 are
tightly  bound from  below by  the requirement  that $X  < 2.5  \times
10^{-6}$, as displayed in  Fig.~\ref{Fig:abundancias}.  We also assert
that the same  evolution leading the system to the  observed region of
$P_{orb}- M_{2}$  plane produces  a high-mass  pulsar as  indicated by
dedicated                        observations                       by
\citet{2012ApJ...760L..36R}. Figs.~\ref{Fig:periodos}-\ref{Fig:radios}
display the evolution of the system at a glance.

We have  finally checked that  a variation  of the initial  donor mass
$M_{2}$ within  the whole interval  leading to  the BW track  does not
lead  to  dramatic  changes  in   the  initial  periods,  and  that  a
bifurcation {\it locus}  can be calculated quite  accurately.  Work in
progress  devoted  to  the  whole  population  of  ``redbacks'',  wide
binaries and BW millisecond pulsar systems \citep{2012arXiv1210.6903R}
is underway and will be published elsewhere.

\begin{centering}
\begin{table*}
\caption{\label{Tabla:datos} 
Some  relevant characteristics  of the  binary systems when  they
reach the observed orbital  period for {\it second} time. We
list the initial period, the age, the donor and NS masses, the central
temperature and  density of  the donor star,  the orbital  period, the
surface H abundance, the radius of  the Roche lobe, the radius and the
mean density of the donor star.  For the case of initial periods above
bifurcation we list these quantities at the age of 13~Gyr.}
\begin{minipage}{18 cm} 
\centering
\begin{tabular}{cccccccccccc}
\hline
\hline
$P_{i}$ & Age & $M_{d}$ & $M_{NS}$ & $\log{T_{c}}$ & $\log{\rho_{c}}$ & $P_{orb}$ & $X$ & $R_{L}$ & $R_{*}$ & $\log{\bar{\rho}_{d}}$ \\
$[d]$  & [Gyr] &[$10^{-3}$~\msol]  & [\msol]  & [K] & [$g\; cm^{-3}$] & [$10^{-2}$~d] & & [$10^{-2}$~\rsol]  & [$10^{-2}$~\rsol]  & [$g\; cm^{-3}$]
\\
\hline
0.75 & 6.615 & 15.6 & 2.22 & 6.185 & 2.631 &  6.45 & 0.099     & 8.16 & 6.62 & 76.2 \\
0.80 & 6.270 & 11.5 & 2.22 & 5.880 & 2.671 &  6.45 & 0.056     & 7.42 & 5.71 & 87.5 \\
0.85 & 5.804 & 9.02 & 2.22 & 5.394 & 2.676 &  6.45 & 0.011     & 6.83 & 5.18 & 91.3 \\
0.86 & 6.195 & 9.08 & 2.29 & 5.357 & 2.704 &  6.45 & 0.002     & 6.79 & 5.08 & 97.6 \\
0.87 & 6.432 & 8.99 & 2.29 & 5.339 & 2.706 &  6.45 & $9 \times 10^{-6}$ & 6.76 & 5.05 & 98.4 \\
0.88\footnote{Largest initial period below bifurcation} & 10.08 & 8.98 & 2.28 & 5.333 & 2.705 &  6.45 & 0.00      & 6.76 & 5.05 & 98.3 \\
0.89\footnote{Smallest initial period above bifurcation} & 13.00 & 157  & 2.23 & 7.037 & 4.996 &  36.5 & 0.241     & 53.8 & 3.43 & 5500 \\
0.90 & 13.00 & 175  & 2.22 & 6.986 & 5.119 &  72.4 & 0.254     & 87.6 & 2.93 & 9747 \\
\hline \hline
\end{tabular} 
\end{minipage}
\end{table*} 
\end{centering}


J.E.H. has been supported by Fapesp (S\~ao Paulo, Brazil) and CNPq, 
Brazil funding agencies.

\bsp 

\label{lastpage}

\end{document}